\documentstyle[aps,preprint]{revtex}
\begin{document}
\draft
\preprint{U.C. Irvine TR-98-15}
\title{
Joint Bayesian Treatment of Poisson and Gaussian Experiments in a 
Chi-squared Statistic}
\author{Dennis Silverman}
\address{
Department of Physics and Astronomy \\
University of California, Irvine \\
4129 Frederick Reines Hall \\ 
Irvine, CA 92697-4575} 
\date{\today}
\maketitle

\begin{abstract}
Bayesian Poisson probability distributions for $\bar{n}$ can be 
analytically converted
into equivalent chi-squared distributions.  These can then be combined
with other Gaussian or Bayesian Poisson distributions to make a total 
chi-squared distribution.  This allows the usual treatment of chi-squared
contours but now with both Poisson and Gaussian statistics
experiments.  This is illustrated with the case of neutrino oscillations.
\end{abstract}

\newpage
\section{Introduction}
In analyzing the joint probability for mutual experimental results or
for parameters, often a number of Poisson statistics experiments with
a low number of events may be mixed with Gaussian experiments with
high numbers of events.  It is desirable to combine both types in a
way to maintain the simplicity of a chi-squared distribution for all
of the experiments.  In this paper we show a simple mathematical
identity between the Bayesian Poisson distribution for the average and
an associated chi-squared distribution that allows us to accomplish
this.  We then apply this to the case of neutrino oscillation 
experiments with few events requiring a Poisson treatment and
find the form of the addition to $\chi^2$ from the Poisson 
experiments to combine with Gaussian treated experiments to form
a combined $\chi^2$ to study the oscillation and mixing parameters.
Having achieved the general result of including Poisson experiments
with Gaussian experiments, we then solve the simplest analytical
cases for linear parameter dependences in the appendices.

In section 2 we review the method for joining two chi-squared
distributions into a joint chi-squared distribution.  In section 3 we
review using Bayes' theorem to find the Bayesian Poisson distribution
for the average.  In section 4 we show the exact equivalence of the Bayesian
Poisson distribution for the average to a chi-squared distribution.
We also show the domain of accuracy when a background is present.
In section 5 we derive the joint probability distribution for
combining a single Bayesian Poisson distribution for the average with
a chi-squared distribution.  In section 6 we then use the results of
section 2 to combine in general the Bayesian Poisson distributions for
averages with chi-squared distributions from Gaussian distributions.
In section 7 we apply the method to the analysis of neutrino oscillation
experiments with small numbers of events.  In section 8 we present our
conclusions.

Several appendices complete the necessary tools with expanded
probability tables.  Others solve the simplest analytic cases for
contributions linear in the physical parameters.
Appendix A reviews the comparison of the integrated probability of the
Bayesian Poisson distribution for the average with the classical
Poisson sum which is often used.  Appendix B gives a table of
two-sided confidence level limits for the Bayesian Poisson average for a single
experiment.  Appendix C gives a table of chi-squared confidence levels
which are useful for the joint distribution.  Appendix D gives the
solution for the minimum chi-squared for the case that the means only
depend linearly on the parameters in both the Poisson and Gaussian
distributions.  Appendix E gives the most probable value and limits
for a single linear parameter in the combination of one Poisson
experiment with one Gaussian experiment.  Appendix F examines the consistency
of converting Poisson to chi-squared distributions in the case of combining
two Poisson distributions whose averages depend on one linear parameter.
  
\section{Method of Joining Two Chi-squared Distributions}
First we show the result that will allow us to join Poisson 
distributions for the averages when we relate them to chi-squared
distributions.  We show that the chi-squared distributions
convolute to form a joint chi-squared distribution.

The basic chi-squared distribution with N degrees of freedom is
\begin{equation}
f_N(\chi^2) = \frac{(\chi^2)^{\frac{N}{2}-1} e^{-\frac{\chi^2}{2}} }
{2^{\frac{N}{2}} \Gamma(\frac{N}{2}) },
\end{equation}
with norm
\begin{equation}
1 = \int_0^\infty d \chi^2 f_N (\chi^2).
\end{equation}

The convolution integral for combining two chi-squared distributions
for $N_1$ and $N_2$ to produce a joint chi-squared distribution is
\begin{equation}
f_N(\chi^2) = \int_0^{\chi^2} d \chi_1^2 f_{N_1} (\chi_1^2)
f_{N_2} (\chi^2 - \chi_1^2)
\label{merge}
\end{equation}
where $\chi_2^2$ is replaced by $(\chi^2-\chi_1^2)$.
By substituting chi-squared distributions in the above, and changing variable
the integration variable to $t=\chi_1^2/\chi^2$, we get
\begin{equation}
f_N(\chi^2)=\frac{1}{2^{(N_1+N_2)/2} \Gamma(\frac{N_1}{2}) 
\Gamma(\frac{N_2}{2})}
e^{-\chi^2/2} (\chi^2)^{(N_1+N_2)/2-1}
\int_0^1 dt~ t^{N_1/2-1} (1-t)^{N_2/2-1}.
\end{equation}
Using the formula for the $t$ integral, which is a beta
function equal to $\Gamma(N_1/2) \Gamma(N_2/2) / \Gamma((N_1+N_2)/2)$,
one sees that the result $f_N(\chi^2)$ 
is the chi-squared distribution function for $N = N_1 + N_2$.
(The analogous formula for joining two Poisson distributions, with
averages $\bar{n}_1$ and $\bar{n}_2$ to produce $n_t$ total events is
\begin{equation}
P(n_t;\bar{n}_t) = \sum_{n_1=0}^{n_t} P(n_1;\bar{n}_1) P(n_t-n_1;\bar{n}_2),
\end{equation}
where $\bar{n}_t = \bar{n}_1 + \bar{n}_2$.)

\section{Poisson Distribution and Bayes Theorem for Limiting 
\boldmath{$\bar{n}$}} According to Bayes'
Theorem\cite{general,agostini,feldman}, the probability for a given
``theoretical parameter average'' $\bar{n}$ given an observed number
of events $n$, $P(\bar{n}; n)$, is proportional to the probability of
observing $n$ events from a Poisson distribution with an average
number of events $\bar{n}$, or $P(n; \bar{n})$\cite{baker}.  The
latter is
\begin{equation}
P(n; \bar{n}) = \frac{\bar{n}^n e^{-\bar{n}} }{n !}.
\end{equation}
The probability distribution for $\bar{n}$, $P(\bar{n}; n)$, is 
proportional to this\cite{weight}, subject to the normalization condition that
the probability for all possible $\bar{n}$ should integrate to unity
\begin{equation}
\int_0^\infty d \bar{n} P(\bar{n}; n) = 1.
\label{norm}
\end{equation}
This is satisfied by the formula for $P(n; \bar{n})$ without further 
renormalization, since the integral is seen to be the form
for $\Gamma(n+1)/n! = 1$.  Thus we have the normalized distribution
for $\bar{n}$ which we call the Bayesian Poisson distribution for the 
average\cite{rainwater}.
\begin{equation}
P(\bar{n}; n) = \frac{\bar{n}^n e^{-\bar{n}} }{n !}.
\label{poisson}
\end{equation}

\section{Connection of the Bayesian Poisson Distribution for the Average to 
a Chi-squared Distribution}

We will show a mapping of the variables
$(\bar{n}, ~n)$ from a Bayesian Poisson distribution for the average
to $(\chi^2, N)$ for a chi-squared distribution that keeps the
identical probability distribution and integration of the Poisson
distribution, but is now in a chis-squared form.  This may be used by
itself using usual chi-squared probabilities and contours, or included
with other chi-squared joined experiments by the convolution
integral in section 2.

The chi-squared distribution to be integrated over $d\chi^2$ for $N$
degrees of freedom is
\begin{equation}
f_N(\chi^2) = \frac{1}{2 \Gamma(N/2)} e^{-\chi^2/2} 
\left(\frac{\chi^2}{2}\right)^{N/2-1}.
\label{chis}
\end{equation}   
This is identical to the $\bar{n}$ distribution to be integrated over
$\bar{n}$
\begin{equation}
P(\bar{n}; n) = \frac{1}{n!} e^{-\bar{n}} \bar{n}^n
\end{equation}
with the identification of 
\begin{equation}
\bar{n} = \frac{\chi^2}{2}, \quad {\rm or} \quad \chi^2 = 2 \bar{n},
\end{equation}
and 
\begin{equation}
n = N/2 - 1, \quad {\rm or} \quad N = 2(n+1).
\end{equation}
The equivalency of the two forms is noted in the Particle Data Group
article on statistics\cite{pdgequiv}, but they do not use it to merge
experiments into a chi-squared distribution. 
The identity includes the integrals over ranges of probabilities in
$\bar{n}$ or equivalently in $\chi^2$ using
\begin{equation}
d\bar{n} = \frac{1}{2} d \chi^2
\end{equation}
Thus a Poisson with n events now counts mathematically as a chi-squared
distribution with $N = 2 n + 2$ degrees of freedom.

If the prior probability is of a logarithmic, 
power law preserving form preferred by statisticians,
$P(\bar{n})=1/\bar{n}$, then the normalized Bayesian Poisson distribution
for the average is directly seen to be the same as that for the uniform prior 
for $n-1$ events, $P(\bar{n};n-1)$\cite{general}.  Since the Poisson form
was the only requirement for the above connection between Bayesian Poisson and 
chi-squared distributions, the results still hold for the logarithmic
prior, but with $n$ replaced by $n-1$, so that $N_{P-log}= 2 n$.

For cases with an unknown mean signal number of events $\bar{n}_S$ plus
an exact known background average $\bar{B}$, the Bayesian Poisson distribution
for the mean $(\bar{n}_S+\bar{B})$ when $n_T$ events are observed 
is\cite{helene84}
\begin{equation}
P(\bar{n};n_T)=
\frac{(\bar{n}_S+\bar{B})^{n_T} e^{-(\bar{n}_S+\bar{B})}}
{\Gamma(n_T+1,\bar{B})},
\end{equation}
where $\Gamma(n_T+1,\bar{B})$ is the incomplete Gamma function. This results
from the normalization over only non-negative values of $\bar{n}_S$.
However, this factor could ruin the simple convolution properties on which this
paper is based.  In cases where $\bar{B}$ is small and $n_T$ a few events, this
correction is small and the $\Gamma(n_T+1,\bar{B})$ can be replaced by
$n_T!$ with little error, and the simple formulas of this paper can again
be used with $n=n_T$ and $\bar{n}=\bar{n}_T=\bar{n}_S+\bar{B}$. To see when
this occurs we note that 
\begin{equation}
\Gamma(n_T+1,\bar{B}) = n_T! \left(1 + \bar{B} + \frac{\bar{B}^2}{2!} 
+\ldots+ \frac{\bar{B}^{n_T}}{n_T!}\right) e^{-\bar{B}}.
\end{equation}
For small $\bar{B}$ the above correction factor to $n_T!$ has leading term
$(1-\bar{B}^{n_T+1}/(n_T+1)!)$, giving hope of its being small if $n_T$ is
not very small and $\bar{B}$ is.  One way to state this is to give the
value of $\bar{B}$ for each $n_T$ at which the correction factor becomes
a given value.  The following Table I gives the values at which the correction
factor becomes 5\% and 1\%.

\begin{tabular}{|c||c|c|c|c|c|c|c|c|c|c|c|}
\hline
\multicolumn{12}{|c|}{Table I: $\bar{B}$ Limits}\\
\hline \hline
$n_T$: & 0 & 1 & 2 & 3 & 4 & 5 & 6 & 7 & 8 & 9 & 10 \\\hline
5\%   & 0.05 & 0.36 & 0.82 & 1.37 & 1.97 & 2.61 & 3.29 & 3.98 & 4.70 & 5.43 &
   6.17 \cr
1\%   & 0.01 & 0.15 & 0.44 & 0.82 & 1.28 & 1.79 & 2.33 & 2.91 & 3.51 & 4.13 &
   4.77 \cr
\hline
\end{tabular}
 
\section{Derivation of Joint Probability for a Bayesian Poisson Distribution 
for the Average and a Chi-squared Distribution}

Here we demonstrate the derivation of the product probability for the
case of one Poisson distribution for the average with a chi-squared 
distribution for $\chi^2_G$ with $N_G$ degrees of freedom
formed either from Gaussians or from joint Gaussian and Poisson 
distributions.
The integrated product probability is
\begin{equation}
1 = \int_0^\infty d \bar{n} P(\bar{n};n)
\int_0^\infty d \chi^2_G f_{N_G}(\chi^2_G)
\end{equation}
We convert the integral over the average $\bar{n}$ to the variable
$\chi^2_P = 2 \bar{n}$ and rewrite using section 4
\begin{equation}
d\bar{n}P(\bar{n};n)=f_{N_P}(\chi^2_P) d\chi^2_P
\end{equation}
with $N_P=2 n+2$.
Into the new integral we now introduce the total $\chi^2$ by inserting
$1=\int_0^\infty d\chi^2 \delta(\chi^2-\chi^2_P-\chi^2_G)$ and use this to
do the $d\chi^2_G$ integral, which limits $\chi^2_P \leq \chi^2$ and gives
\begin{equation}
1= \int_0^\infty d\chi^2 \int_0^{\chi^2} d\chi^2_P f_{N_P}(\chi^2_P)
f_{N_G}(\chi^2-\chi^2_P).
\end{equation}
By the chi-squared convolution integral, the second integral is 
$f_{N_P+N_G}(\chi^2)$, which is the resultant probability distribution
for this case, with $\chi^2=\chi^2_P+\chi^2_G$. The result is an
exact joint chi-squared probability combining a Poisson experiment
with a chi-squared distribution from previously combined experiments.

\section{Merging Bayesian Poisson and Chi-squared Distributions}
Now that we have a $f_N(\chi^2)$ distribution Eq.(\ref{chis}) that is
equivalent to a Bayesian Poisson parameter distribution in value and in its
probability integral, we can merge this (independent of its origin) 
with other chi-squared
distributions using Eq.~(\ref{merge}), the convolution, to
obtain the final $\chi^2$ distribution.

The results can now be used, for example, in finding $\chi^2$ contours
corresponding to various confidence levels.  We must remember that a
single Poisson experiment 
with a uniform prior now counts as $N = 2 (n+1)$ degrees of
freedom, where $n$ is the number of observed events in the Poisson
distribution.  While this sounds counter-intuitive, we recall
that the form of the $\chi^2$ distribution that we are using also has
$\chi^2$ replaced by $2 \bar{n}$, and with the above replacements,
$\chi^2$ per degree of freedom $N$ or $\chi^2/N = 2 \bar{n}/(2(n+1))$,
approaches $1$ at large $n$ since
$n$ is within $\sqrt{\bar{n}}$ of $\bar{n}$.

If $M_P$ is the number of Poisson experiments with $n_i$ events in the
$i$'th experiment, we associate with each $N_i = 2 n_i + 2$ degrees of
freedom.  We call the associated theoretical Poisson averages $\bar{n}_i$.
The total Poisson degrees of freedom becomes
\begin{equation}
N_P = \sum_{i=1}^{M_P}N_i = \sum_{i=1}^{M_P}(2 n_i + 2).
\end{equation}
With the alternate choice of a logarithmic prior, $N_{P-log}=
\sum_{i=1}^{M_P} 2 n_i$.
We now convolute the Poisson distributions for the average in the
chi-squared forms, Eqs.~(8-17) with the chi-squared distribution of
$N_G$ Gaussian experimental 
degrees of freedom which have a chi-squared $\chi_G^2$.  
The result will use the joint chi-square
\begin{equation}
\chi^2_{PG} = 2 \sum_{i=1}^{M_P} \bar{n}_i + \chi^2_G.
\end{equation}
From successive convolutions in Eq.~(\ref{merge}), the combined chi-squared
distribution for the Poisson plus Gaussian distributions is finally
\begin{equation}
f_{(N_G + N_P)} (\chi^2_{PG}).
\end{equation}

We emphasize that these results are an exact treatment, not 
involving large $n$ or other approximations.  As in the standard treatment,
if $N_{\rm par}$ is the number of parameters that are being fitted, then
the number of degrees of freedom is $dof=N = N_G+N_P-N_{\rm par}$.
In Appendix B we show how the $\chi^2$ limits 
at various confidence levels for two-sided distributions are related to Poisson
sums.
In Appendix B we give an expanded Table II that can be used for 
two-sided $\chi^2$ limits at
given confidence levels.  In Appendix C an expanded table for single-sided
$\chi^2$ values or $\chi^2$ contours for $N$ up to 25 corresponding
to various confidence levels.  In the respective appendices we also 
give Mathematica programs
to be used for larger $N$ or other confidence levels.

This method has been applied in analyzing the constraints of many experiments 
on new flavor changing neutral current models of CP violation in $B$ meson
decay asymmetries\cite{silverman}.  There, all experiments have a Gaussian
distribution, except for an experiment\cite{ktopi} where one event has been 
seen in $K^+ \to \pi^+ \nu \bar{\nu}$ and is treated with an additional
$\chi^2_P = 2 \bar{n}$ and adding four degrees of freedom.  In that case,
$\bar{n}$ is a function of the down quark mixing matrix elements as are
the other experiments. That analysis also provides an example of the
sensitivity to the choice of a uniform or logarithmic prior probability
distribution.  With the uniform prior, the total number of degrees of freedom
is seven, and the chi-squared limits are at 8.2, 12.0, and 14.3 for 
1-$\sigma$, 90\% (1.64-$\sigma$), and 2-$\sigma$ confidence levels, 
respectively.  With the
logarithmic prior, the total number of degrees of freedom decreases by two
to five, and the chi-squared limits are at 5.89, 9.24, and 11.3, for the
1-$\sigma$, 90\%, and 2-$\sigma$ confidence levels, respectively.  The 
chi-squared per degree of freedom ratios stay withing 10\% of each other
between the two cases.  However, use of the logarithmic prior does move the
contours in by two to three units or about 1/2 of a standard deviation, and 
thus gives tighter bounds.

Parenthetically we add that in the limit of large $n$ and $\bar{n}$,
just as the Poisson distribution becomes a Gaussian, so does the
equivalent chi-squared distribution.  The chi-squared distribution in
Eq.~(\ref{chis}) becomes
\begin{equation}
G(\xi,\sigma) = \frac
{e^{-\xi^2/2 \sigma^2}}{\sqrt{2\pi}\sigma},
\end{equation}
where in our variables $\sigma=\bar{n}^{1/2}=(\chi^2/2)^{1/2}$,
$\xi=n-\bar{n}=(N/2-1)-\chi^2/2$, and $d\chi^2 = -2 d\xi$.  Since
$|\xi|$ is confined to the order of $\sigma$ for large $n$ and
$\bar{n}$, the difference $|N/2-1-\chi^2/2|$ is confined to the order
of $\sqrt{\chi^2/2}$ or $\sqrt{N/2}$ for large $N$ and $\chi^2$.

\section{Neutrino Oscillation Experiments}
Here we shall see that using the combined Poisson method for small
numbers of events per bin leads to a result which considers only
the total number of events in a single Poisson distribution, and 
makes the two methods identical.

\subsection{Appearance Neutrino Oscillation Experiments}
For example, we consider a $\nu_\mu \to \nu_e$ appearance experiment.  
Let $n_i^0$ be the number of expected $\nu_\mu$ in the i'th bin at
energy $E_i$, $n_i$ the number of observed events in that bin, and 
$b_i$ the known background in that bin. With the two neutrino oscillation
formula, the average number of electrons in that bin will be
\begin{equation}
\bar{n}_i = n_i^0 \sin^2{(2 \theta)} \sin^2{(1.27 \delta m^2 L/E_i)}.
\end{equation}

By the method of expressing Bayesian Poisson's in the chi-squared 
formalism, we get the total chi-squared as a linear sum of expected
events for each bin from Eq.~14, if the $b_i$ are sufficiently small
\begin{equation}
\chi^2 = \sum_i (2 \bar{n_i}+ 2 b_i).
\end{equation}
Also, the number of degrees of freedom is twice the total number of 
observed events $n$ when using the logarithmic prior
\begin{equation}
N_{P-log} = 2 \sum_i n_i \equiv 2 n.
\end{equation}
The sum of background events is denoted by $B = \sum_i b_i$.

With small bin size $\Delta E_i$, $n_i^0 = (dn/dE) \Delta E_i$, and
the sum of the expected number of events at full mixing can
be converted into an integral 
\begin{equation}
n^0(\delta m^2) \equiv \int dE \frac{dn}{dE} 
\sin^2{(\frac{1.27 \delta m^2 L}{E})}.
\end{equation}
So we now have a binning independent form for $\chi^2$ from the sum over
bins
\begin{equation}
\chi^2=2 \sin^2{(2 \theta)} n_0(\delta m^2)+ 2 B
\end{equation}
and a binning independent number of degrees of freedom $N=2n$.
The probability distribution is now
\begin{equation}
f_N(\chi^2) = f_{2n}(2 \sin^2{(2 \theta)} n_0(\delta m^2)+ 2 B).
\end{equation}
We set 90\% CL limits using a one-sided CL if there is no signal, 
and a two-sided CL if there is a signal. For the one-sided CL limit,
the average background $\bar{B}$ has to be less than or equal to
$0.05$ events for the $n_T=0$ Poisson to be accurately normalized.

Going backwards from a chi-squared distribution to its equivalent
Poisson distribution, this chi-squared result is equivalent to a Bayesian
Poisson distribution for the average with
\begin{equation}
\bar{n} = \chi^2/2 =  \sin^2{(2 \theta)} n_0(\delta m^2)+  B,
\end{equation}
with $n$ events observed.  This is the same as the usual approach of
grouping all events into one bin of the total number of events,
which is used if there are few events.  As in the
case of Eq. 14, B must be small enough not to significantly affect the
normalization.

\subsection{Disappearance Neutrino Oscillation Experiments}
For a disappearance experiment, the expected number of events
per bin is
\begin{equation}
\bar{n}_i = n_i^0(1 - \sin^2{(2 \theta)} \sin^2{(1.27 \delta m^2 L/E_i)}).
\end{equation} 
Using the same sums as for the appearance experiment, and defining
the sum of the coefficients of the 1 term or the total number of expected
neutrino events without oscillation as
\begin{equation}
n^0 = \int dE \frac{dn}{dE},
\end{equation}
we have the probability distribution
\begin{equation}
f_N(\chi^2) = f_{2n}(2 n^0 -2 \sin^2{(2\theta)} n^0(\delta m^2)+2B).
\end{equation}

If the total number of events is large enough to use a Gaussian
approximation, these are then the same results as using a single
Gaussian in the usual method for comparing the total number of events
with and without oscillation.
But even with a limited total number of events, the formulas above
with a chi-squared distribution 
are an improvement over a Gaussian, as long as the
background $b_i$ are small enough in each bin.

\subsection{General Comments on Oscillation Results}
What we have achieved is that for a small number of events, we have
found the $\chi^2_P$ for the Poisson neutrino appearance and
disappearance experiments, Eqs. (27) and (32), respectively, that 
can be added to $\chi^2$ from other
Poisson or Gaussian neutrino experiments to determine neutrino
oscillation and mixing parameters using standard $\chi^2$ methods with
$2n$ extra degrees of freedom as in the logarithmic prior case.
The drawback is that the result is equivalent to a comprehensive bin 
in energy containing all events.  When the number of particles per
each energy bin becomes significant, it is better to use a Gaussian
for each bin to derive information contained in the detailed energy
spectrum.

\subsection{One-Sided Chi-squared Limits on Oscillation}

We find a contour in the $(\sin^2{(2\theta)},\delta m^2)$ plane
where for the probability distribution $f_{2n}(\chi^2)$ the 
amount of probability contained in the major part is the confidence
level CL.  The appropriate one-sided chi-squared limits $\chi^2_{CL^+}(2n)$
for $n$ observed events and $N_P=2(n+1)$ for a uniform prior or 
$N_{P-log}=2n$ for a logarithmic prior are found in Table III of 
Appendix C.

\subsubsection{Appearance Experiment}

In practice, for each $\delta m^2$ we find the value of $\sin^2{(2\theta)}$
such that the bound becomes an equality
\begin{equation}
2 \sin^2{(2 \theta)} n_0(\delta m^2)+2B \leq \chi^2_{CL^+}(2n).
\end{equation}
$CL^+$ means that for a 95\% CL limit, only 5\% is left off of the upper
part of the distribution. The excluded region is where the left-hand-side
is larger than the chi-squared upper CL limit, giving an upper bound
on $\sin^2{(2\theta)}$.

\subsubsection{Disappearance Experiment}
Here, the excluded region is where the left-hand-side is smaller
than the chi-squared lower CL limit, and the allowed region is
\begin{equation}
2 n^0 -2 \sin^2{(2\theta)} n^0(\delta m^2) +2B \geq \chi^2_{CL^-}(2n),
\end{equation}
which again restricts the result with an upper bound on  $\sin^2{(2\theta)}$.

\subsection{Large {\boldmath $n$} Gaussian Approximation}
While the previous results were accurate for small $b_i$, for large $n$ 
we may use the approximation that the chi-squared distribution
resembles a Gaussian distribution near its peak
\begin{equation}
f_{2n}(\chi^2) \approx \frac{1}{\sqrt{2 \pi n}}
\exp{(-\frac{(n-\chi^2/2)^2}{2n})}.
\end{equation}
A one-sided 95\% CL limit which leaves 5\% on one side, is at the same
deviation (in $\chi^2/2$) from the center of the Gaussian as the usual
two-sided 90\% CL limit which leaves 5\% on both sides.  This occurs at
\begin{equation}
|n-\chi^2/2| = 1.64 \sigma = 1.64 \sqrt{n}.
\end{equation}

This yields the chi-squared limits below.
Since the multiplier term of $\sin^2{(2\theta)}$ can average to
a half or be less than that, values of $\sin^2{(2\theta)}$ greater
than one can be reached in these limits, and they must be cut off at
one.

For appearance experiments, the two sided 90\% CL limits are
\begin{equation}
\sin^2{(2\theta)} \leq \frac{n+ 1.64\sqrt{n}-B}{n^0(\delta m^2)}.
\end{equation}

For disappearance experiments, the two sided 90\% CL limits are
\begin{equation}
\sin^2{(2\theta)} \leq \frac{n^0+B-n+1.64\sqrt{n}}{n^0(\delta m^2)}.
\end{equation}

For one-sided 90\% CL limits we use $1.28\sigma$.

\section{Conclusions}
In conclusion, we have shown how the simplicity of the Bayesian
$\chi^2$ analysis can be exactly extended to include experiments with
a small number of events which are described by a Bayesian Poisson
distribution for the average.  This precise analytic treatment (provided
the background is small) is useful since it uses the simple chi-squared
treatment for all experiments, even if some experiments have too few
events to be a standard Gaussian.  We have provided useful tables
for the method by extending them to larger $n$ to accompany the
larger number of degrees of freedom used.  We have analyzed neutrino
oscillation experiments and showed how the analytic combination of
Poisson bins through the equivalent chi-squared distributions leads to
the standard Poisson result for the total number of events.  However,
using the equivalence to a chi-squared distribution, we have found
the appropriate $\chi_P^2$ to add to the $\chi^2$ from other experiments
to use standard $\chi^2$ methods.

\acknowledgements
 
The author thanks Peter Meyers, Bill Molzon, and Jonas Schultz for
helpful discussions.  This research was supported in part by the
U. S. Department of Energy under Contract No. DE-FG0391ER40679.

\appendix
\section{Comparison With Other Formulas Used For Poisson Parameter
Limits}
For completeness we include here some properties of and a comparison between
the classical (or frequentist) and Bayesian Poisson limits on $\bar{n}$.  
The methods are
given full discussion by R. D. Cousins in Ref.~1.
The classical Poisson parameter distribution used for the upper $\bar{n}$
limit is to sum the Poisson distributions
$P(n;\bar{n})$
from $n+1$ events to infinity, when the number of observed events is $n$, 
and use it as the probability for $\bar{n}$ when $\bar{n}$ is greater
than $n$.
We show that the Bayesian Poisson parameter distribution Eq.~(\ref{poisson})
integrated from zero to a cutoff $n_c$ agrees with the above 
formulation\cite{general}.  
First we do the integrated probability for $\bar{n}$ from
$n_c$ to infinity by integrating $e^{-\bar{n}}$ by parts  
\begin{eqnarray}
I(n_c;n) &=& \int^{\infty}_{n_c} d\bar{n} \frac{\bar{n}^n}{n!} e^{-\bar{n}} \\
         &=& \left[ \frac{\bar{n}^n}{n!}(-e^{-\bar{n}})\right]^\infty_{n_c}
 + \int^\infty_{n_c} d\bar{n} 
 \frac{\bar{n}^{n-1}}{(n-1)!} e^{-\bar{n}} \\
         &=& P(n;n_c) + I(n_c;n-1).
\label{integral}
\end{eqnarray}
Continued integration by parts shows that the integral over a semi-infinite 
interval beginning at $n_c$ of the Bayesian Poisson parameter distribution 
is\cite{general,abram}
\begin{equation}
I(n_c;n) = P(n;n_c) + P(n-1;n_c) + \ldots + P(0;n_c).
\end{equation}
The two methods are now seen to be equivalent using $\bar{n} = n_c$ and
the fact that the Poisson terms sum\cite{general} to 1
\begin{eqnarray}
\sum_{n'=n+1}^\infty P(n';\bar{n}) &=& 1 - \sum_{n'=0}^n P(n';\bar{n}) \\
                                   &=& 1 - I(\bar{n};n) \\
&=& \int^{n_c}_{0} d\bar{n} \frac{\bar{n}^n}{n!} e^{-\bar{n}}.
\label{poissonlower}
\end{eqnarray}
from Eqs.~(\ref{norm}) and (\ref{integral}).

For $n$ the number of observed events, the rule for the ``1-$\sigma$''
upper limit on $n_c$ is to find $n_c^+$ such that 84\% of the time there would
be greater than $n$ events. Since ``1-$\sigma$'' means 32\% is outside the
central region, 16\% should occur on one side.  Thus the sum from $n+1$ to
infinity is set equal to 0.84
\begin{equation}
\sum_{n'=n+1}^\infty P(n';n_c^+) =
\int^{n_c^+}_{0} d\bar{n} \frac{\bar{n}^n}{n!} e^{-\bar{n}}
= 1 - I(n_c^+,n) = 0.84
\end{equation} 
from Eq.~(A6).  So for the upper ``1-$\sigma$'' limit, $n_c^+$,
both the Bayesian result of setting the integral of the Poisson distribution 
for the average in Eq.(\ref{poissonlower}) 
equal to 0.84 and the sum of higher $n$ agree. 

For the lower 1-$\sigma$, the classical rule of setting the sum from 
0 to $n-1$ 
equal to 0.84 to determine $n_c^-$ (or the sum from $n$ to $\inf$ set to 0.16)
gives
\begin{equation}
\sum_{n'=0}^{n-1} P(n';n_c^-) = I(n_c^-;n-1) = 0.84.
\end{equation}
This is not the same as setting the integral of the 
Bayesian Poisson distribution 
for the average from 0 to $n_c^-$ equal to 0.16
\begin{equation}
1-I(n_c^-;n) = \int^{n_c^-}_{0} d\bar{n} \frac{\bar{n}^n}{n!} e^{-\bar{n}} =
0.16 \quad {\rm or} \quad I(n_c^-;n)=0.84
\end{equation} 
from Eq.~(\ref{poissonlower}).
To see the difference, we note from Eq.(\ref{integral})
\begin{equation}
I(n_c^-;n) = P(n;n_c^-) + I(n_c^-;n-1).
\end{equation}

With the prior chosen to be $1/\bar{n}$, the lower limits agree but not
the upper\cite{general}.
\section{Table of Bayesian Poisson Central Limits for the Average
and Two-Sided Chi-squared Limits}

The Bayesian Poisson average central interval limits with uniform prior
are the upper or lower $n_c^{\pm}$ limits 
as in Eq.~(A8) or Eq.~(A10) beyond which the confidence level is below a
given value.  This is in analogy with the $\bar{x}\pm\sigma$ one $\sigma$
limits in a single Gaussian distribution, where half of the excluded intervals
on each side are used in the integral limits (0.16 on each side for 1$\sigma$).
The following Table II covers
lower and upper limits out to 3$\sigma$, and for $n=0$ to $n=24$.  

Comparing Eq. A1 with the results of section 4 we have the relation between
the Poisson integral over the average and the equivalent chi-squared 
integral at a given confidence level, say $CL^+$
\begin{equation}
I(n_c^+;n)=\int^{\infty}_{n_c^+} d\bar{n} \frac{\bar{n}^n}{n!} e^{-\bar{n}}
=\int^{\infty}_{(\chi_c^2)^+}d \chi^2 f_N(\chi^2)=CL^+,
\end{equation}
with $(\chi_c^2)^+ = 2 n_c^+$ and $N=2(n+1)$.
For the lower confidence level limits
\begin{equation}
1-I(n_c^-;n)=\int^{n_c^-}_{0} d\bar{n} \frac{\bar{n}^n}{n!} e^{-\bar{n}}
=\int^{(\chi_c^2)^-}_{0}d \chi^2 f_N(\chi^2)=CL^-.
\end{equation}
So in both cases, we can get the $\chi^2$ limits from Table II also by
using
\begin{equation}
(\chi_c^2)^\pm = 2 n_c^\pm.
\end{equation}

Table II was
produced from the following Mathematica program (except for the n column),
which can be used to extend the table as needed. It also shows the
actual confidence levels used for the various column designations
in the program.
\begin{eqnarray*}
 & & {\rm <<Statistics`ContinuousDistributions`} \\
{\rm cl}& =& \{0.0013499,0.01, 0.0227501,0.1,0.158655,0.5,0.841345,0.9,
0.9772500,0.99,0.9996500\} \\
{\rm navgtable} &:=&
  {\rm N[Table[0.5*Quantile[ChiSquareDistribution[k],cl[[i]]],\{k,4,50,2\},
\{i,1,11\}], 4]} \\
 & & {\rm TeXForm[navgtable//TableForm].}
\end{eqnarray*}
For $n=0$ events observed, the one-sided confidence interval upper bounds 
are meaningful as opposed to two-sided intervals.  The upper limits 
of intervals starting from zero which contain 0.6827, 0.90, 0.95,
0.9545, 0.99, and 0.9973 probability are 1.15, 2.30, 3.00, 3.09, 4.61, and
5.9, respectively.  G. J. Feldman and R. D. Cousins use an approach which 
carefully covers both single and double-sided cases\cite{feldman}.

\newpage
\begin{tabular}{|c||c|c|c|c|c|c|c|c|c|c|c|}
\hline 
\multicolumn{12}{|c|}{Table II:  Bayesian Poisson Central 
Limits for the Averages $n_c^-$ and $n_c^+$} \\ \hline \hline
n & -3$\sigma$ & 0.01 & -2$\sigma$ & 0.1 & -1$\sigma$ & 0.5
& 1$\sigma$ & 0.9 & 2$\sigma$ & 0.99 & 3$\sigma$ \\\hline
   1 & 0.05288 & 0.1486 & 0.2301 & 0.5318 & 0.7082 & 1.678 & 
   3.300 & 3.890 & 5.683 & 6.638 & 10.39 \cr 
   2 & 0.2117 & 0.4360 & 0.5963 & 1.102 & 1.367 & 2.674 &
   4.638 & 5.322 & 7.348 & 8.406 & 12.47 \cr 
   3 & 0.4653 & 0.8232 & 1.058 & 1.745 & 2.086 & 3.672 & 
   5.918 & 6.681 & 8.902 & 10.05 & 14.38 \cr 
   4 & 0.7919 & 1.279 & 1.583 & 2.433 & 2.840 & 4.671 & 
   7.163 & 7.994 & 10.39 & 11.60 & 16.18 \cr 
   5 & 1.175 & 1.785 & 2.153 & 3.152 & 3.620 & 5.670 & 
   8.382 & 9.275 & 11.82 & 13.11 & 17.90 \cr 
   6 & 1.603 & 2.330 & 2.758 & 3.895 & 4.419 & 6.670 &
   9.584 & 10.53 & 13.22 & 14.57 & 19.56 \cr 
   7 & 2.068 & 2.906 & 3.391 & 4.656 & 5.232 & 7.669 &
   10.77 & 11.77 & 14.59 & 16.00 & 21.17 \cr 
   8 & 2.563 & 3.507 & 4.046 & 5.432 & 6.057 & 8.669 &
   11.95 & 12.99 & 15.94 & 17.40 & 22.75 \cr 
   9 & 3.084 & 4.130 & 4.719 & 6.221 & 6.891 & 9.669 & 
   13.11 & 14.21 & 17.27 & 18.78 & 24.30 \cr
   10 & 3.628 & 4.771 & 5.409 & 7.021 & 7.734 & 10.67 & 
   14.27 & 15.41 & 18.58 & 20.14 & 25.82 \cr 
   11 & 4.191 & 5.428 & 6.113 & 7.829 & 8.585 & 11.67 & 
   15.42 & 16.60 & 19.87 & 21.49 & 27.32 \cr 
   12 & 4.772 & 6.099 & 6.828 & 8.646 & 9.441 & 12.67 & 
   16.56 & 17.78 & 21.16 & 22.82 & 28.80 \cr 
   13 & 5.367 & 6.782 & 7.555 & 9.470 & 10.30 & 13.67 & 
   17.70 & 18.96 & 22.43 & 24.14 & 30.26 \cr 
   14 & 5.977 & 7.477 & 8.291 & 10.30 & 11.17 & 14.67 & 
   18.83 & 20.13 & 23.70 & 25.45 & 31.70 \cr 
   15 & 6.599 & 8.181 & 9.036 & 11.14 & 12.04 & 15.67 &
   19.96 & 21.29 & 24.95 & 26.74 & 33.13 \cr 
   16 & 7.233 & 8.895 & 9.789 & 11.98 & 12.92 & 16.67 &
   21.08 & 22.45 & 26.20 & 28.03 & 34.55 \cr 
   17 & 7.877 & 9.616 & 10.55 & 12.82 & 13.80 & 17.67 &
   22.20 & 23.61 & 27.44 & 29.31 & 35.95 \cr 
   18 & 8.530 & 10.35 & 11.32 & 13.67 & 14.68 & 18.67 & 
   23.32 & 24.76 & 28.68 & 30.58 & 37.34 \cr 
   19 & 9.193 & 11.08 & 12.09 & 14.53 & 15.57 & 19.67 &
   24.44 & 25.90 & 29.90 & 31.85 & 38.72 \cr 
   20 & 9.863 & 11.83 & 12.87 & 15.38 & 16.45 & 20.67 &
   25.55 & 27.05 & 31.13 & 33.10 & 40.10 \cr 
   21 & 10.54 & 12.57 & 13.65 & 16.24 & 17.35 & 21.67 & 
   26.66 & 28.18 & 32.34 & 34.35 & 41.46 \cr 
   22 & 11.23 & 13.33 & 14.44 & 17.11 & 18.24 & 22.67 &
   27.76 & 29.32 & 33.55 & 35.60 & 42.82 \cr 
   23 & 11.92 & 14.09 & 15.23 & 17.97 & 19.14 & 23.67 &
   28.87 & 30.45 & 34.76 & 36.84 & 44.17 \cr 
   24 & 12.62 & 14.85 & 16.03 & 18.84 & 20.03 & 24.67 &
   29.97 & 31.58 & 35.96 & 38.08 & 45.51 \cr  
\hline
\end{tabular}
\newpage

\section{Table for Chi-squared Values at Various
Confidence Levels}
Since the joint method for $n$ events requires $\chi^2$ for 
$N = 2(n+1)+N_G-N_{par}$ for a uniform prior, or for
$N = 2n +N_G-N_{par}$ for a logarithmic prior, both of which can be large, 
we give here a table of chi-squared values for various confidence levels for
large $N$ up to 25, and a program with which one can generate further limits.

In the following table, N is the number of degrees of freedom, and
the designations of
1, 2, and 3 $\sigma$ correspond to 1-CL of 0.682689, 0.954500, and
0.997300, respectively.  The Mathematica program used to generate the
table is
\begin{eqnarray*}
 & & {\rm <<Statistics`ContinuousDistributions`} \\
{\rm cl}      &=& \{0.682689, 0.9, 0.954500,0.99,0.997300\} \\
{\rm cstable} &:=& {\rm N[Table[Quantile[ChiSquareDistribution[k],cl[[i]]]
, \{k,1,25\},\{i,1,5\}],4]}\\
         & & {\rm TeXForm[cstable // TableForm]}.
\end{eqnarray*}

\newpage
\begin{tabular}{|c||c|c|c|c|c|}
\hline
\multicolumn{6}{|c|}{Table III:  Chi-squared Limits} \\
\hline \hline
   N  & 1$\sigma$ & 0.90 & 2$\sigma$ & 0.99 & 3$\sigma$ \\\hline
   1  & 1.000 & 2.706 & 4.000 & 6.635 & 9.000 \cr
   2  & 2.296 & 4.605 & 6.180 & 9.210 & 11.83 \cr 
   3  & 3.527 & 6.251 & 8.025 & 11.34 & 14.16 \cr 
   4  & 4.719 & 7.779 & 9.716 & 13.28 & 16.25 \cr 
   5  & 5.888 & 9.236 & 11.31 & 15.09 & 18.21 \cr 
   6  & 7.038 & 10.64 & 12.85 & 16.81 & 20.06 \cr 
   7  & 8.176 & 12.02 & 14.34 & 18.48 & 21.85 \cr 
   8  & 9.304 & 13.36 & 15.79 & 20.09 & 23.57 \cr 
   9  & 10.42 & 14.68 & 17.21 & 21.67 & 25.26 \cr 
   10 & 11.54 & 15.99 & 18.61 & 23.21 & 26.90 \cr 
   11 & 12.64 & 17.28 & 19.99 & 24.72 & 28.51 \cr 
   12 & 13.74 & 18.55 & 21.35 & 26.22 & 30.10 \cr
   13 & 14.84 & 19.81 & 22.69 & 27.69 & 31.66 \cr 
   14 & 15.94 & 21.06 & 24.03 & 29.14 & 33.20 \cr 
   15 & 17.03 & 22.31 & 25.34 & 30.58 & 34.71 \cr 
   16 & 18.11 & 23.54 & 26.65 & 32.00 & 36.22 \cr 
   17 & 19.20 & 24.77 & 27.95 & 33.41 & 37.70 \cr
   18 & 20.28 & 25.99 & 29.24 & 34.81 & 39.17 \cr 
   19 & 21.36 & 27.20 & 30.52 & 36.19 & 40.63 \cr 
   20 & 22.44 & 28.41 & 31.80 & 37.57 & 42.08 \cr
   21 & 23.51 & 29.62 & 33.07 & 38.93 & 43.52 \cr 
   22 & 24.59 & 30.81 & 34.33 & 40.29 & 44.94 \cr 
   23 & 25.66 & 32.01 & 35.58 & 41.64 & 46.36 \cr 
   24 & 26.73 & 33.20 & 36.83 & 42.98 & 47.76 \cr 
   25 & 27.80 & 34.38 & 38.07 & 44.31 & 49.16 \cr  
\hline
\end{tabular}

\newpage
\section{Solution for Chi-squared Expansion About the Minimum
for the Linear Parameter Dependence Case}

For the case where the theoretical values for the mean in the Gaussian and
Poisson distributions are linear in parameters to be fitted, 
the minimum of $\chi^2$ and its quadratic expansion about the 
minimum can be 
found analytically using the same method as for pure Gaussian 
distributions\cite{mathews,pdglinear}.  While this may prove useful,
in the usage here, however, the maximal probability of the $\chi^2$
distribution is not at the
minimum $\chi^2$, but at $\chi^2 \approx n$.

In the method of expressing Poisson distributions for the average as
$\chi^2$ distributions in this paper, the final $\chi_{GP}^2$ is
\begin{eqnarray}
\chi^2_G &=& \sum_{i=1}^{N_G} 
\frac{(y_i-F_i(\mathbf{\alpha}))^2}{\sigma_i^2}, 
\quad {\rm and} \\
\chi^2_{GP} &=& \chi^2_G + 2 \sum_{\ell=1}^{N_P} 
\bar{n}_\ell( \mathbf{\alpha}),
\label{chisgp}
\end{eqnarray}
where $\mathbf{\alpha}$ is the set of $k$ parameters $\alpha_m$.  The 
experiments described by $(y_i,F_i)$ can even be totally different, and
the $F_i$ and $\bar{n}_\ell$ are assumed to be linearly expandable in the 
parameters $\alpha_m$
\begin{eqnarray}
F_i(\mathbf{\alpha}) &=& \sum_{n=1}^k \alpha_n f_{in}, \quad {\rm and}\\
\bar{n}_\ell(\mathbf{\alpha})&=&\sum_{j=1}^k n_{\ell j} \alpha_j.
\end{eqnarray}
Minimizing $\chi^2_{GP}$ with respect to each $\alpha_m$ gives rise to
the vector ${\bf g}$ and matrix $V^{-1}$ with components
\begin{eqnarray}
g_m &=& \sum_{i=1}^{N_G} y_i \frac{f_{im}}{\sigma_i^2}
    - \sum_{\ell=1}^{N_P} n_{\ell m}, \quad {\rm and} \\
V^{-1}_{mn} &=& \sum_{i=1}^{N_G} \frac{f_{in} f_{im}}{\sigma_i^2}.
\end{eqnarray}
Using the inverse matrix $V$, the values of the parameters that give the
minimum $\chi^2_{GP}$ are given by
\begin{equation}
\hat{\alpha} = V \mathbf{g},
\end{equation}
with the effect of the $\bar{n}_k$ terms entering through ${\bf g}$.
The minimum value of $\chi^2_{GP}$ is Eq.~(\ref{chisgp}) evaluated at
$\alpha = \hat{\alpha}$.
$\chi^2_{GP}$ can then be rewritten in terms of $\alpha$ away
from the minimum values as
\begin{equation}
\chi^2_{GP} = \chi^2_{GP-{\rm min}} + 
(\alpha - \hat{\alpha})^T V^{-1} 
(\alpha - \hat{\alpha}).
\end{equation}

\section{Solution of One Bayesian Poisson Distribution with One
Gaussian Distribution and One Linear Parameter}
We present here the solution for the single linear parameter case
with one Bayesian Poisson and one Gaussian distribution.  For the unknown
parameter $a$, we have the theoretical relations $\bar{n} = a c_P$ 
for the Poisson average, and
$\bar{x} = a c$ with known standard deviation $\sigma$ for the
Gaussian average, where coefficients $c_P$ and $c$ are given, and
$n$ and $x$ are the results of the respective experiments.
Then
\begin{equation}
\chi^2_{PG} = 2 \bar{n} + (x-\bar{x})^2/\sigma^2
\end{equation}
With one parameter to be fitted, the number of joint degrees of
freedom with the equivalent chi-squared method with a uniform prior
is $N=2 n + 2 +1-1 = 2n+2$
where one degree of freedom is cancelled by the one parameter.
For the logarithmic prior, $N=2n+1-1=2n$, which gives tighter
$\chi^2$ limits.

The minimum of $\chi^2_{PG}$ occurs at
\begin{equation}
\bar{a}c = x - \sigma^2 c_P/c
\end{equation}
giving the minimum chi-squared
\begin{equation}
\chi^2_{min} = 2 x (c_P/c) - \sigma^2 (c_P/c)^2.
\end{equation}

When $\chi^2_{PG}$ is set equal to a certain upper limit boundary at
$\chi^2_{lim}$, there are bounds on the range of $a$ given by
\begin{equation}
a^\pm_{lim} = \bar{a} \pm \frac{\sigma}{c}\sqrt{\chi^2_{lim}-\chi^2_{min}}.
\end{equation}

For physical reasons we may want $\bar{a}$ to be positive when
$c_P$ and $c$ are positive.  Looking at $\bar{x} = \bar{a}c$ above,
we see that $\bar{x}$ and $\bar{a}$ are positive when
$x \bar{x}/\sigma^2 \geq \bar{n}$.  In order to use a Gaussian,
we expect at least a 3-$\sigma$ separation of the peak from zero, or
$x/\sigma \geq 3$ and $\bar{x}/\sigma \geq 3$.  Thus for 
$\bar{n} \leq 9$, this method works and $\bar{a} \geq 0$.
For $\bar{n} \geq 9$, $\bar{n}/\sqrt{\bar{n}} \geq 3$ and we 
can start using a Gaussian instead of a Poisson for the 
$\bar{n}$ experiment.  The same reasoning follows through if for
example we require a 5-$\sigma$ separation from zero to use a 
Gaussian.

\section{Two Poisson Distributions With One Linear Parameter}

We approach this problem both from Bayes theorem directly, and from
converting the Bayesian Poisson distributions to chi-squared distributions
as proposed in this paper.  For the latter we then merge the chi-squared 
distributions to a single chi-square distribution for the linear parameter 
and then convert that back to a joint Poisson distribution, to compare
to the direct approach.  For the case of the logarithmic prior we
find consistency.

The averages of the experiments are theoretically given by the parameter
$a$ with respective known coefficients $\bar{n}_1 = a c_1$ and 
$\bar{n}_2 = a c_2$.
The direct Bayesian result is proportional to the probability for 
observing the experimental values $n_1$ and $n_2$ given a value of $a$
\begin{eqnarray}
{\rm Prob}(a;n_1,n_2) &=& P(n_1;ac_1) P(n_2;ac_2) P(a)/(P(n_1)P(n_2)) \cr
&\propto& (a c_1)^{n_1} (a c_2)^{n_2} e^{-a c_1} e^{-a c_2} P(a) \cr
&\propto& (a(c_1+c_2))^{(n_1+n_2)} e^{-a(c_1+c_2)} P(a) \cr
&\propto& P(a(c_1+c_2);n_1+n_2) P(a).
\end{eqnarray}
For the uniform prior, $P(a)=1$, the normalized result is 
$P(a(c_1+c_2); n_1+n_2)$, integrating over $da (c_1+c_2)$.
For the logarithmic prior with $P(a)=1/a$, the normalized result is the same
as the uniform prior with total $n$ lowered by 1, or 
$P(a(c_1+c_2); n_1+n_2-1)$, integrating over $da (c_1+c_2)$.

If we now start with the method in this paper, we take the joint Bayesian
result as the product of the Bayesian Poisson for each experiment as
if they were independent, $P(ac_1;n_1) P(ac_2;n_2)$ times either the 
uniform prior $d\bar{n}_1 d\bar{n}_2$ or the logarithmic 
prior $d\bar{n}_1 d\bar{n}_2
/(\bar{n}_1 \bar{n}_2)$.  The logarithmic prior is equivalent to 
$P(ac_1;n_1-1)P(ac_2;n_2-1)$
with a uniform prior.  Converting the uniform case to chi-squared distributions
gives the convolution of the product $f_{2n_1+2}(2ac_1) f_{2n_2+2}(2ac_2)$
leading to $f_{2n_1+2n_2+4}(2ac_1+2ac_2)$.  Converting this back to a
Poisson distribution for the average gives $P(ac_1+ac_2;n_1+n_2+1)$ for the
uniform prior, which is inconsistent with the direct uniform Bayesian result 
in the 
previous paragraph.  For the logarithmic prior, converting to chi-squared
distributions gives the convolution of the product 
$f_{2n_1}(2ac_1) f_{2n_2}(2ac_2)$ which is $f_{2n_1+2n_2}(2ac_1+2ac_2)$.
Converting this back to a Poisson distribution for the average gives
\begin{equation}
P(ac_1+ac_2;n_1+n_2-1) \propto P(ac_1+ac_2;n_1+n_2)
\frac{da(c_1+c_2)}{a(c_1+c_2)},
\end{equation}
which is consistent with the direct Bayesian result
for the logarithmic prior in the previous paragraph.

In the combined form as a single Bayesian Poisson distribution for the
average, both upper and lower limits on $a$ for a given central confidence 
interval can be found using the table in appendix B.  The case where no
events were observed in either experiment can also be dealt with using
one-sided bounds, which are also given in appendix B.


\begin{references}
\bibitem{weight} Bayes' Theorem also contains the ratio of the a priori
probabilities of $P(\bar{n})/P(n)$.  We take both as unity, or 
$P(\bar{n})=1$, which is called a uniform prior.
\end{references}
\end{document}